\documentclass[useAMS,usenatbib]{mn2e}

\usepackage{graphicx}

\title[Core and jet in PG~1700$+$518]{The radio core and jet in the broad absorption line quasar PG~1700$+$518}
\author[Yang et al.]{J.\,Yang$^{1}\thanks{E-mail:yang@jive.nl}$,
F.~Wu$^{2,3}$,
Z.~Paragi$^{1,4}$ and
T.~An$^{2,5,6}$
\\
\\
$^{1}$Joint Institute for VLBI in Europe, Postbus~2,  7990\,AA Dwingeloo, The Netherlands \\
$^{2}$Shanghai Astronomical Observatory, CAS, 200030 Shanghai, P.R. China \\
$^{3}$Graduate University of the Chinese Academy of Sciences, 100049 Beijing, P.R. China \\
$^{4}$MTA Research Group for Physical Geodesy and Geodynamics, POB\,91, H-1521 Budapest, Hungary \\
$^{5}$Key Laboratory of Radio Astronomy, Chinese Academy of Sciences, P.R. China \\
$^{6}$Netherlands Institute for Radio Astronomy (ASTRON), Postbus~2, 7990\,AA Dwingeloo, The Netherlands
}

\begin{document}

\date{Accepted 2011 Oct. Received 2011 Oct. ; in original form 2011 Oct. xx}

\pagerange{\pageref{firstpage}--\pageref{lastpage}} \pubyear{2012}
\maketitle
\label{firstpage}

\begin{abstract}
The blue-shifted broad absorption lines (BAL) or troughs are observed in Active Galactic Nuclei (AGNs) when our line of sight is intercepted by a high speed outflow (wind), likely originating in the accretion disc. The outflow or wind can shed light on the internal structure obscured by the AGN torus. Recently, it has been shown that this outflow is rotating in the BAL quasar PG~1700$+$518, further supporting the accretion disc origin of the wind. With the purpose of giving independent constraints on the wind geometry, we performed high-resolution European VLBI Network (EVN) observations at 1.6~GHz in 2010. Combining the VLBI (Very Long Baseline Interferometry) results with the Very Large Array (VLA) archival data at 8.4~GHz, we present its jet structure on scales of parsec (pc) to kiloparsec (kpc) for the first time. The source shows two distinct jet features in East-West direction with a separation of around 4~kpc. The Eastern feature, which has so far been assumed to hide the core, is a kpc-scale hot spot, which is completely resolved out in the EVN image. In the western jet feature, we find a compact jet component, which pinpoints the position of the central black hole in the galaxy. Jet components on both sides of the core are additionally detected in the Northwest-Southeast direction, and they show a symmetric morphology on scale of $<$1~kpc. This two-sided jet feature is not common in the known BAL quasars and indicates that the jet axis is far away from the line of sight. Furthermore, it is nearly parallel to the scattering plane revealed earlier by optical polarimetry. By analogy to polar-scattered Seyfert\,1 galaxies, we conclude that the jet likely has a viewing angle around 45$\degr$. The analogy is further supported by the recent report of significant cold absorption in the soft X-rays, a nearly unique feature to polar-scattered Seyfert galaxies. Finally, our observations have confirmed the earlier finding that the majority of radio emission in this galaxy arises from AGN activity rather than star-formation.

\end{abstract}


\section{Introduction}
\label{sec1}
Broad absorption line (BAL) quasars constitute a significant fraction, $\sim$15 percent, in the entire quasar population \citep[e.g.][]{tru06}. It is generally believed that their broad absorption lines are an indication of a high-velocity outflow or wind \citep[e.g.][]{axo08}. In the BAL quasar PG~1700$+$518 (J1701$+$5149) it was found by \citet{you07} that the broad Balmer lines show not only radial motion but also significant rotation ($\sim$4000~km\,s$^{-1}$), indicating that the outflow is related to a rotating accretion disc wind. The geometry of this outflow thus is of high interest because it may shed light on the innermost structure of active galactic nucleus
(AGN) itself in this system.

Assuming that both the accretion disc and the torus have the same axis as the jet in PG~1700$+$518, high resolution radio observations of the jet can provide constraints on the wind geometry. The jet axis is perpendicular to the polarisation position angle of optical continuum emission in most Seyfert\,2 galaxies and parallel in around three quarter of Seyfert\,1 galaxies \citep[e.g.][]{ant83, smi04}. These polarisation properties can be well explained within the model of two scattering regions \citep{smi04,axo08}. The scattered emission is dominated by polar scattering in Seyfert\,2 galaxies, and by equatorial scattering in Seyfert\,1 galaxies. The transition between Type 1 and 2 Seyfert galaxies is polar-scattered Seyfert\,1 galaxies. In this case our line of sight is right along the edge of the torus, thus the polarized light from the equatorial scattering region is sufficiently suppressed, while the broad wings of the Balmer lines are not affected.

The quasar PG~1700$+$518 was first discovered by \citet{sch83} in an optical survey and later identified by \citet{wam85} as a rare, nearby BAL quasar. It is an ultra-bright ($m_\mathrm{R}\sim15.5$) infrared source, while its host galaxy is one of the most molecular gas-rich Palomar-Green QSO hosts. The star-forming molecular gas mass was estimated to be $\sim6\times10^{10}M_\odot$ from CO(1$\rightarrow$0) emission observations \citep{eva09}. A ring-shaped companion galaxy was found in deep infrared observations \citep{sto98} at 1.6~arcsec North of the quasar with a similar redshift ($z\sim0.29$), indicating a merging process between the two galaxies.

The radio counterpart of PG~1700$+$512 is weak. It has a total flux density of $20\pm1$~mJy at 1.4 GHz \citep[e.g.][]{bar96}. The source is not clearly resolved with the Very Large Array (VLA) below $\sim$5~GHz. At higher frequencies the source shows two extended emission regions, separated by $\sim$1~arcsec (4~kpc) in the East-West direction \citep{hut92, kuk98, kel94}. VLA images at 8.4~GHz \citep{kuk98} reveal that the western emission region extends towards North-West. \citet{blu98} reported a possible detection in the western emission region with the Very Long Baseline Array (VLBA) at 8.4~GHz, but only at the 3$\sigma$ level and with no information on the jet axis.

The letter is organised in the following sequence. We introduce our observations and data reduction in Section~\ref{sec2} and then present the imaging results in Section~\ref{sec3}. Based on these high-quality images, we identify the radio core, present the case that the quasar is a polar-scattered Seyfert\,1-like object, and discuss the discrepancy between the radio and infrared-derived star-formation rates in
Section~\ref{sec4}. Finally, we summarise our results in Section~\ref{sec5}. Throughout this paper, we used the cosmology corrected scale: 4.2~pc~mas$^{-1}$ at $z=0.29$, derived from the following cosmology:~$H_0=73$~km\,s$^{-1}$\,Mpc$^{-1}$, $\Omega_\mathrm{M}=0.27$ and $\Omega_\mathrm{\Lambda}=0.73$.

\section{Observations and Data Reduction}
\label{sec2}

\subsection{The 1.6~GHz EVN observations}
\label{sec2-1}

The Very Long Baseline Interferometry (VLBI) observations of PG~1700$+$518 were performed at 1.6~GHz with the European VLBI Network (EVN) on 2010 November 5 (project code: EY012). There were nine stations (Effelsberg, Westerbork in the phase-array mode, Jodrell Bank (MkII), Onsala, Medicina, Torun, Svetloe, Zelenchukskaya, Badary) participating in the experiment. Since PG~1700$+$518 is not a bright radio source, we adopted the phase-referencing technique with a cycle time of 6 minutes: 2 minutes on the calibrator and 4 minutes on the target. The reference calibrator J1701$+$5133 is quite close (15 arcmin) to our target and has a compact and bright core in the VLBA Imaging and Polarisation Survey \citep[VIPS,][]{hel07}. The adopted J2000 coordinate \citep{pet11} of the reference source is $\mathrm{RA}=17^\mathrm{h}01^\mathrm{m}22\fs34991$, $\mathrm{Dec.}=+51\degr33\arcmin49\farcs6890$. The EVN observations lasted for 6 hours totally, and used a recording rate of 512~Mbps (dual polarisation, 16 channels, 8 MHz per channel, 2-bit sampling). Correlation was done with the EVN MkIV Data Processor at JIVE (Joint Institute for VLBI in Europe) with an integration time of 2 seconds and 16 frequency points per channel.

Following the online EVN data reduction guide, the \emph{a-priori} calibration was done in AIPS \citep[Astronomical Image Processing System;][]{gre03}. The correlation amplitudes were calibrated using the measured gain curves and system temperatures for most of the telescopes; we used nominal SEFDs (System Equivalent Flux Density) for Badary, Jodrell Bank, Svetlo and Zelenchukskaya. Fringe-fitting was done in two steps: first we removed the instrumental inter-channel delay using a two-minute scan of the bright calibrator J1638$+$5720 (``manual phasecal''), and then we solved for all the delay, delay-rate and phase solutions of calibrators via global fringe-fitting of the whole experiment. The phase solutions of PG~1700$+$518 came from linear interpolation of calibrator solutions. Bandpass calibration was done using J1638$+$5720. Finally, the data were averaged in each subband and split into single-source files.

Self-calibration and imaging were done in the Caltech software package Difmap \citep{she94}. The phase-reference source J1701$+$5133 was imaged first. It has a total flux density of 55~mJy, dominated by a compact core (92\% of the total flux density). The final amplitude and phase self-calibration was done in AIPS using the calibrator image obtained in Difmap. Both the amplitude and the phase solutions were applied to the target source. There was no self-calibration applied during the imaging of PG~1700$+$518. Furthermore, we did circular Gaussian model fitting in Difmap \citep{she94} to determine the characteristic parameters of each discrete emission region. The model-fitting results are summarised in Table~\ref{tab1}.

\subsection{The 8.4~GHz VLA archive data}
\label{sec2-2}

We analysed a VLA data set (project code: AB0512), which was available in the NRAO Data Archive. The source PG~1700$+$518 was observed with the VLA in configuration A at 8.4~GHz for $\sim$2.3~hours on 1988 December 15. The absolute flux density was calibrated with 3C~286. The nearby point source 1739$+$522 was used to further calibrate the amplitude and phase. The final imaging was done without any self-calibration in Difmap \citep{she94}.

\begin{table}
\caption{The results of the circular Gaussian model fitting for the jet features in PG~1700$+$518. Note that
component C (core) is the reference origin and the last column is total flux density.} \label{tab1}
\scriptsize
\setlength{\tabcolsep}{3.7pt}
\centering
\begin{tabular}{rccrrrc}
\hline
  % after \\: \hline or \cline{col1-col2} \cline{col3-col4} ...
Array & Freq. & Comp. &  Radius   & P.A.~~        &   Size            & Flux              \\
      & (GHz) &       & (mas)     & ($^\circ$)~~~~&        (mas)      & (mJy)             \\
\hline
VLA   & 8.4   & C     &   $0\pm0$ &               &        $<$30      &  $0.87\pm0.09$    \\
VLA   & 8.4   & NW    & $215\pm3$ & $-32.2\pm0.8$ &    $134\pm3$      &  $0.44\pm0.05$    \\
VLA   & 8.4   & SE    & $228\pm4$ & $141.1\pm1.0$ &    $139\pm4$      &  $0.35\pm0.04$    \\
\vspace{0.5em}
VLA   & 8.4   & E     & $912\pm3$ &  $93.6\pm0.2$ &    $203\pm3$      &  $1.69\pm0.17$    \\
EVN   & 1.6   & C     &   $0\pm0$ &               &      $3\pm1$      &  $1.9\pm0.2$      \\
EVN   & 1.6   & NW    & $231\pm2$ & $-39.0\pm0.6$ &        $>$40      &  $>$1.6         \\
EVN   & 1.6   & SE    & $270\pm2$ & $133.3\pm0.6$ &        $>$40      &  $>$1.5         \\
\hline
\end{tabular}
\end{table}

\section{EVN and VLA Imaging Results}
\label{sec3}

\begin{figure*}
  % Requires \usepackage{graphicx}
  \includegraphics[width=\textwidth]{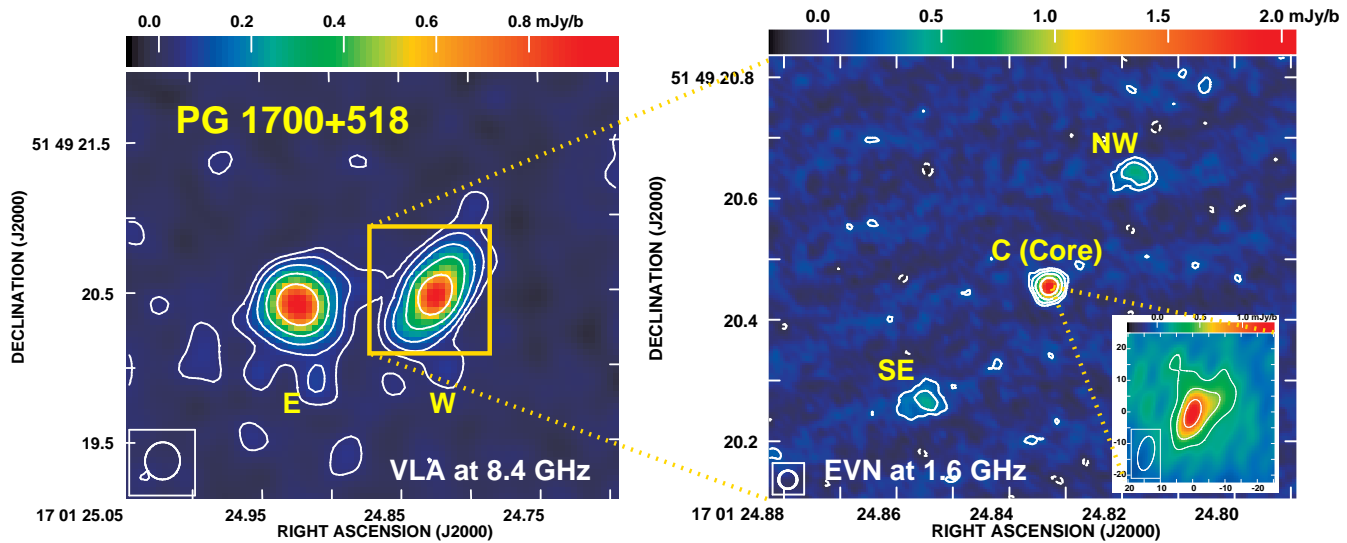}\\
\caption{The VLA and EVN intensity images of PG~1700$+$518. The golden rectangle in the left VLA image shows the field of the right EVN image. The bottom-right inset panel in the EVN image exhibits the inner substructure of component C (core). The contours start from 3$\sigma$ off-source noise level (VLA:~0.036~mJy\,beam$^{-1}$, EVN:~0.15~mJy\,beam$^{-1}$) and increase by a factor of 2.}\label{fig1}
\end{figure*}

The final EVN and VLA imaging results are collected in Fig.~\ref{fig1}. The left VLA image shows that PG~1700$+$518 has two jet features, marked as E and W and separated by 912$\pm$3~mas, in agreement with the earlier imaging results \citep{hut92, kel94, kuk98}. Compared with their images, our image has the better resolution: a nearly circular beam with a size of 0.25 arcsec. The eastern feature has a peak brightness of 0.99~mJy\,beam$^{-1}$, slightly higher than the western one. In addition, there is a hint of surrounding diffuse emission \citep[cf.][]{hut92} at the 4$\sigma$ level (1$\sigma$=0.012 mJy\,beam$^{-1}$). The western jet feature is elongated along the SE-NW direction \citep[cf.][]{kuk98}. With three circular Gaussian components (SE, C and NW), the elongated structure can be well fitted.

The right panel in Fig.~\ref{fig1} is the EVN image which was restored with a 30~mas circular Gaussian, to present the faint, resolved components better (indicated by SE, C, and NW). The inset shows the full resolution map of the brightest feature. Note that the components SE and NW are too extended to be seen with a beam of $<$10~mas, and their total flux density and size in Table~\ref{tab1} are underestimated due to the absence of short baselines ($<$200~km) in the array. One might suspect that these quasi-symmetric components are simply result of amplitude calibration errors in our data (dominated by errors on the Ef-Wb baseline). A slight calibration error may arise from the fact that the pointing centre of the VLBI observations was taken from the NASA Extragalactic Database (NED), and it is located about halfway between the eastern and the western VLA components. But this $\sim$0.5 arcsecond offset is still small compared to the phased array WSRT beam of 10 arcseconds at 1.6~GHz. We further note that SE and NW are not exactly symmetric around C, and their orientation agrees with both the elongation of the western VLA component, as well as the elongation of the resolved central feature C. However, improved $uv$-coverage on $\sim$100\,km baselines is necessary to map better the diffuse emission in this source. In particular, we see a hint of emission in the VLBI data near the eastern VLA component as well, but we cannot reliably image that extended feature since it is almost completely resolved out on VLBI baselines.

The absolute position of the brightest and most compact component C with respect to the assumed J(2000) coordinates of the reference source are $\mathrm{RA}=17^\mathrm{h}01^\mathrm{m}24\fs82640$, $\mathrm{Dec.}=+51\degr49\arcmin20\farcs4473$, with an error of about 1~mas. The measured position of component C is about 50 mas offset from the 3$\sigma$ detection at 8.4~GHz claimed by \citet{blu98}, and this difference is significant even when we consider the large uncertainties and the fact that different calibrators were used in the two experiments.

The brightness temperature can be estimated with \citep[e.g.][]{kel88}:
\begin{equation} \label{eq2}
    T_\mathrm{B}=1.22\times10^{12}(1+z)\frac{S_\nu}{\theta^2\nu^2_\mathrm{obs}}
\end{equation}
where $S_\nu$ is the measured flux density in Jy, $\nu_\mathrm{obs}$ is the observing frequency in GHz and $\theta$ is the angular size in mas. Component C has a brightness temperature of $1\times$10$^{8}$~K, significantly brighter than any other jet components ($\sim$10$^{6}$~K).

\section{Discussion}
\label{sec4}

\subsection{Identification of the radio core}
\label{sec4-1}

Earlier it was believed that the radio core was likely hiding in the eastern VLA feature rather than the western one because of its more compact radio morphology on the VLA scales \citep[e.g.][]{kel94}. The optical position was not accurate enough to exclude the association \citep{kuk98}. With our new EVN data this possibility can be firmly excluded since component E is significantly resolved. The nature of component E is not clear. A possible explanation is a residual hot spot in the fading lobe, considering the surrounding diffuse emission.

Another possible location of the radio core is in the western jet feature. Both the VLA and EVN images reveal an elongated jet. The central component C exhibits much more compact morphology and higher brightness temperature than any other components. When compared with the VLA flux density $\sim$0.8~mJy\,beam$^{-1}$ at 8.4~GHz and assuming no large variability over 22 years, it has a spectral index of $\sim-0.5$ ($S_\nu\propto\nu^{\alpha}$). This is less steep than the total flux density spectrum: $\alpha\sim-1$ \citep{bar96}. If it indeed has a steep spectrum,  this is not surprising in radio-weak AGNs, in particular Seyfert nuclei \citep[e.g.][]{ori10, kha10}. Compared with the later VLA observations with the same configuration by \citet{kuk98} on 1992 December 11, the total flux density is quite stable in component E, while, a factor of 1.7 lower in component W. Such large change is likely associated with the central AGN activity. Note that this variation is too large ($>$10\%) to be explained as a result of the higher phase errors when the phase self-calibration is not possible. The central compact component is also close to the X-ray counter part detected recently by \emph{XMM-Newton} \citep{bal11} at $\mathrm{RA}=17^\mathrm{h}01^\mathrm{m}24\fs8$, $\mathrm{Dec.}= +51\degr49\arcmin20\arcsec$ (J2000) and \emph{Chandra} (ObsID 11853) at $\mathrm{RA}=17^\mathrm{h}01^\mathrm{m}24\fs7$, $\mathrm{Dec.}= +51\degr49\arcmin21\farcs1$ (J2000). The smoothed \emph{Chandra} X-ray image shows a hint on a jet-like extension, spatially coincident with the jet structure observed in component W. Therefore, the component C is the radio core, indicating the location of the AGN.

The radio core has a brightness temperature slightly lower than other VLBI-detected BAL quasars \citep[e.g. 10$^9$~--~10$^{10}$~K,][]{liu08}, but it is at the proper level considering the possible free-free absorption at low frequencies and the large jet viewing angle (see Sect.~\ref{sec4-2}). While the overall spectral index of C is still steep (i.e. not showing the inverted spectrum expected for free-free absorption), one should carefully align multi-frequency images at high resolution in the future, to properly image the spectral index distribution within the core. The large difference between the VLBA \citep{blu98} and our current EVN measurements indicates that either the 3$\sigma$ VLBA detection was an artifact, or there is indeed a large offset between the two frequencies, and the real radio core is heavily free-free absorbed at 1.6~GHz.

The AGN origin of component C is further strengthened by its high radio luminosity ($1.9\times10^{39}$~erg\,s$^{-1}$) which is $\sim$100~times higher than the brightest supernovae and SNR. We further note that the ratio of the core radio luminosity to the X-ray luminiosity of this source \citep[2.5$\times10^{42}$~erg\,s$^{-1}$;][]{bal11} is $L_{R}/L_{X}=0.76\times10^{-3}$, significantly higher than the $L_{R}/L_{X}\sim10^{-5}$ ratios observed in radio quiet quasars \citep{laor08}.

The quasar shows a strong and broad Mg\,II absorption line at ejection velocities between 7000 and 18\,000~km\,s$^{-1}$ \citep{pet85}. Assuming a viewing angle of $45\degr$ (c.f. Sect.~\ref{sec4-2}), we can give a lower limit of the jet speed: $0.1c$, where $c$ is light speed. As the core has a low brightness temperature and the two-sided jet components have similar flux density, size and angular separation from the core, the jet is most likely non-relativistic, as found in most Seyfert galaxies \citep{mid04}.

It is pointed out by \citet{eva09} that PG~1700$+$518 may have significant jet activity as the star-formation rate derived from radio luminosity is a factor of 13 higher than that from the infrared luminosity. This explanation is now confirmed in view of our finding of a radio core with two-sided jet, and a possible large-scale hot spot (component E in the VLA map). The star-forming activity in PG~1700$+$518 is likely boosted by the interaction with the nearby companion galaxy. Moreover, there may be an extra contribution from jet-induced star-formation. There is a bright infrared knot at a radius $\sim$1~arcsec (4.2~kpc) and position angle $\sim$144$^\circ$ \citep{mar03, sto98}. The detection of the knot is significant in the deconvolved H-band image (private communication with I.~M\'arquez). We note that the IR knot (indication of a star-forming region) is right along the jet direction. One may speculate that the jet interacts with the interstellar medium near component SE and then bends toward North-East, as indicated by the extended emission (VLA component E). However at present we have no firm evidence for this scenario and future high sensitivity observations (also in terms of brightness temperature) are needed to confirm the possibility.

\subsection{Similarities with polar-scattered Seyfert\,1 Galaxies}
\label{sec4-2}
The quasar PG~1700$+$518 is a Seyfert\,1-like object within the framework of the unified AGN model since it has broad emission lines. Interestingly, its jet is not as close to the line of sight as in most broad line quasars because the jet components on both sides of the core have been detected with similar total flux density, size and angular separation from the core. This quasi-symmetric jet morphology is typically not found in BAL quasars \citep[e.g.][]{liu08, jia03}. The jet axis, revealed by our observations, is perpendicular to the polarisation position angle \citep[$55\pm4^\circ$, e.g.][]{sch99} of the optical continuum observations. In other words, the scattering plane is along the jet direction. Similar geometry is observed in most Seyfert\,2 galaxies and only in around a quarter of Seyfert\,1 galaxies. This is broadly explained as a result of a polar scattering region above the accretion disc. These polar-scattered Seyfert\,1 galaxies represent the transition between Type 1 and 2 Seyfert galaxies within the model of both polar and equatorial scattering regions \citep{smi04, axo08}. By analogy to the classification of Seyfert galaxies, PG~1700$+$512 has a nucleus resembling polar-scattered Seyfert\,1 galaxies with a jet viewing angle ($\sim$45$^\circ$), comparable with the torus opening angle.

The X-ray observations of polar-scattered Seyfert\,1 galaxies \citep{jim08} have recently uncovered a nearly ubiquitous signature, the presence of warm and/or cold ($N_\mathrm{H}>10^{21}$~cm$^{-2}$) absorption. In the case of PG~1700$+$518, its soft X-ray weakness requires $N_\mathrm{H}>10^{24}$~cm$^{-2}$ \citep{bal11}, higher than in most polar-scattered Seyfert galaxies.

Our classification provides an independent support for the wind geometry inferred from the simulation \citep{you07}. The polarised H$\alpha$ line profile can be well reproduced when the simulation used an intermediate inclination ($\sim$45$^\circ$) and a cylindrical wind geometry rather than other wind geometries, such as a disc-like and rotating equatorial outflow.

% If there is a similar column density of free electrons as well (near the AGN gas ionized), this could well
% result in an optical depth of >>1 at 1.6 GHz for an electron temperature of 10^4 K and using the
% 1 arcsec E-W separation for the size of the free-free absorbing medium. But the alpha~2 inverted spectrum is
% not observed in the core. We addressed this above.

% \subsection{Deviation from the radio-infrared correlation}

\section{Summary}
\label{sec5}
With the new EVN observations and public VLA archive data, we present the jet structure of PG~1700$+$518 on parsec to kilo-parsec scales. We find a compact component with a brightness temperature of $\sim$10$^{8}$~K and identify it as the radio core. The jet components on two sides of the core have similar angular separation from the core, similar size and total flux density, indicating a large viewing angle. The jet axis, observed on scales of $<$1~kpc, is well collimated and perpendicular to the polarisation position angle of the optical continuum emission. Together with the recent report of significant cold absorption in the X-rays, we show that the broad line quasar resembles polar-scattered Seyfert\,1 galaxies, also indicating a viewing angle of $\sim$45$\degr$. Our radio observations confirm the earlier finding that the radio emission is dominated by AGN activity and not by star-formation. Furthermore, the apparent alignment of the jet with a bright infrared feature may hint on the presence of jet-induced star-formation in the galaxy.

\section*{Acknowledgments}
\label{ack}
We thank the anonymous reviewer for his constructive suggestions and Michael Garrett for valuable comments on jet-induced star-formation. T. An thanks the financial support by the Overseas Research Plan for CAS-Sponsored Scholars, the Netherlands Foundation for Sciences (NWO) and the Science \& Technology Commission of Shanghai Municipality (06DZ22101).  F. Wu thanks the JIVE Summer Student Program. The EVN is a joint facility of European, Chinese, South African and other radio astronomy institutes funded by their national research councils. The National Radio Astronomy Observatory is a facility of the National Science Foundation operated under cooperative agreement by Associated Universities, Inc. This research has made use of the NASA/IPAC Extragalactic Database (NED) which is operated by the Jet Propulsion Laboratory, California Institute of Technology, under contract with the National Aeronautics and Space Administration.

\bsp
\label{lastpage}
\end{document}